\documentclass[preprint,aps,prl,psfig]{revtex4}
\usepackage{amsmath}
\usepackage{graphicx}
\usepackage{color}
\newcommand{\be}{\begin{eqnarray}}
\newcommand{\ee}{\end{eqnarray}}
\newcommand{\etal}{{\em et al}}
\newcommand{\jpb}{J. Phys. B }

\begin{document}
 
\title{Evidence for a {\bf T}-shape break-up pattern in the triple photoionization of Li}
 
\author{J. Colgan}
\affiliation{Theoretical Division, Los Alamos National Laboratory,
         Los Alamos, NM 87545}
\author{A. Emmanouilidou}
\affiliation{Department of Physics, Gower Street, University College London, London, UK}
\author{M. S. Pindzola}
\affiliation{Department of Physics, Auburn University, Auburn, AL 36832}

\begin{abstract}

We examine the angular distributions of all three electrons ionized from Li by a single photon
near the triple ionization threshold using a fully quantum-mechanical treatment. 
We find strong evidence for a {\bf T}-shape break-up
pattern at 5~eV excess energy as previously predicted by quasi-classical simulations 
[A. Emmanouilidou and J. M. Rost, \jpb {\bf 39}, 4037 (2006); 
A. Emmanouilidou, P. Wang, and J. M. Rost, \prl {\bf 100}, 063002 (2008)]. This finding is in conflict
with the expected Wannier break-up dynamics of three electrons moving at mutual angles of 120$^{\circ}$,
which is expected to hold at energies a few eV above threshold.
We use our quantum-mechanical approach to explore the physical mechanisms behind this unusual break-up
configuration.
 
\end{abstract}

\maketitle

 
\bigskip

The break-up of atoms near the ionization threshold has been a topic of interest for many years. 
The pioneering work of Wannier \cite{wan53,wan55}, using classical mechanics, revealed the
expected escape dynamics of two electrons moving in the field of an ion (the classic three-body Coulomb problem) as back-to-back emission, 
and led to a threshold law for the energy dependence of single  ionization that was also
applicable to double photoionization. Later work \cite{ks76,ko98,mrb97,mb97} extended this analysis to three outgoing electrons, where it was 
predicted that the three-electron break-up should proceed along the vertices of an equilateral triangle.
The ``triangular" break-up was verified experimentally in $(e,3e)$ coincidence measurements of all three outgoing electrons arising from 
the electron-impact double ionization of He \cite{ren08}. In contrast, the process of triple photoionization of Li,
which leads to the same final state of three electrons moving in the field of a nucleus, has been predicted \cite{er06} to proceed via a
${\bf T}$-shape break-up in the threshold region, that is, two electrons moving back-to-back (anti-parallel) to the third electron which is
at $90^{\circ}$ to this direction. 
The ${\bf T}$-shape break-up was attributed to the different initial state of the Li atom \cite{ewr08} (compared to an electron incident on He), and it was 
shown that the emission of three electrons is, in general, initial state dependent. 

In this work, we use a fully quantum-mechanical approach using the time-dependent close-coupling (TDCC) method to 
probe the triple photoionization of Li and find strong evidence for the prevalence
of a ${\bf T}$-shape break-up at excess energies of 5 eV above the triple ionization threshold. We compare our calculations as far as possible with the 
quasi-classical predictions \cite{er06,ewr08}. Such a comparison can be somewhat involved, for several reasons. 
TDCC calculations become increasingly computationally intensive as the excess energy is lowered, due to the requirement of
using large radial grid sizes and long propagation times to accurately treat the interaction of the slow outgoing electrons. The lowest excess energy we consider
in this work is thus 5~eV, which is (just) a computationally feasible TDCC calculation, and which is a low enough energy at which the ${\bf T}$-shape break-up should be visible,
as predicted by the quasi-classical calculations. 
Also, in the quasi-classical
simulations only the {\it relative} angle between any of the outgoing electrons has any meaning, since there is no reference from which to give an {\it absolute}
angle. In the TDCC calculations (and in any measurement), such a reference is provided by the polarization direction, and the absolute angles of ejection of the
ionized electrons are given with respect to this polarization direction. As we show below, the polarization direction can strongly influence the resulting angular
distributions and therefore, any comparison that is possible between TDCC calculations and the quasi-classical simulations can only be approximate.

In this work, we extend an earlier TDCC approach \cite{cp12} to examine the triple photoionization
of Li at energies close to the triple ionization threshold (at 203.4 eV \cite{nist}).  
The TDCC approach \cite{cpr04,cp12} treats all three electrons equivalently, by expanding
the three-electron Schr\"odinger equation in coupled spherical harmonics, leading to a set of time-dependent coupled partial differential equations that must be solved
for the radial dimensions of all three electrons. 
We present calculations performed at excess energies (E) of 100, 30, 10, and 5~eV and various angular distributions
for the outgoing electrons, and at all possible energy sharings, using a form of the angular distributions described recently \cite{cp12}. 
As the excess energy is lowered, the TDCC calculations become more challenging
in that larger radial meshes and longer propagation times are required. In the calculations reported here, the largest mesh used was $(384)^3$, with 
a spacing of $\Delta r=0.15$ a.u. Test calculations made at smaller mesh spacings of $0.1$ a.u., did not change the resulting electron angular distributions,
even though they resulted in a somewhat more accurate triple ionization threshold. 

In figure~1 we present the angular distributions for triple ionization of Li at four excess energies of 100, 30, 10, and 5 eV. We show the angular distributions
at equal energy sharing between the electrons $E_1=E_2=E_3$ and in the coplanar geometry (so that the plane of ejection of the electrons is in the polarization
plane). Distributions are shown for fixed angles of electrons 1 and 2 at $45^{\circ}$ and $135^{\circ}$, respectively. 
At 100~eV excess energy, the third electron is ejected at an angle of around $270^{\circ}$, which
is the angle that maximizes the third electron's distance from the other two electrons. Smaller lobes are evident at angles near $\sim~240^{\circ}$ and
$\sim~300^{\circ}$. As the excess energy drops to 30~eV, these lobes become more prominent, and move slightly to ejection
angles {\it closer} to the other (fixed) electrons.
As the excess energy drops to 10~eV, these side lobes become the dominant feature of the angular distribution, with the ejection along
$270^{\circ}$ clearly suppressed. At the lowest excess energy considered of 5~eV, the side lobes again move slightly in angle, but are still the dominant
ejection feature. Calculations made at different energy sharings between the outgoing electrons are similar in shape to the distributions presented in figure~1.

\begin{figure}
\label{fig1}
\includegraphics[scale=0.4]{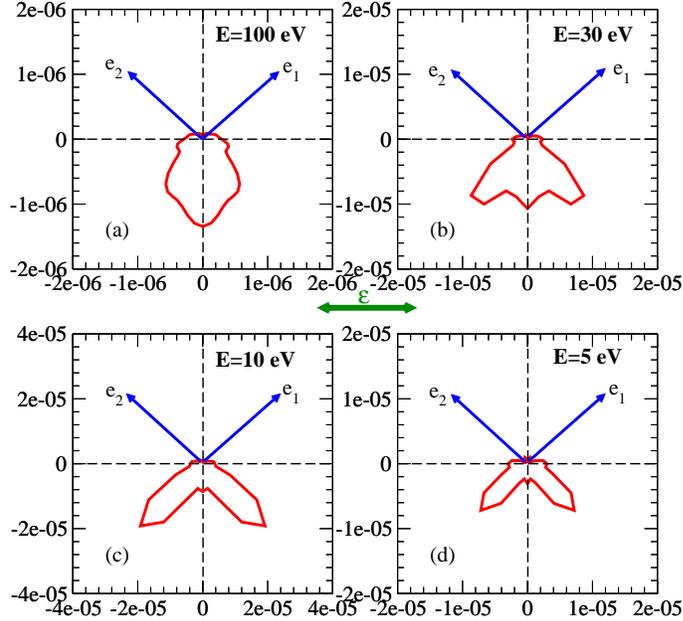}
\caption{(Color online) Pentuple differential cross sections for triple photoionization of Li at 
four excess energies (E) as indicated, $E_1=E_2=E_3$, and for $\theta_1=45^{\circ}$, $\theta_2=135^{\circ}$. Results are presented as a 
function of $\theta_3$ and for coplanar geometry.
All cross sections are in units of b/(sr$^3$ eV$^2$).
}
\end{figure}

The ejection angles made by the third electron at the lower excess energies are seen to be in anti-parallel directions to the fixed electrons 1 and 2. They
thus can be described as a ${\bf T}$-shape that is symmetric about the axis between the fixed electrons. This ${\bf T}$-shape break-up pattern is precisely
that predicted by the earlier quasiclassical calculations \cite{ewr08} and is characterized by two electrons leaving back-to-back, with the remaining electron
perpendicular to this back-to-back axis. The distribution of the third electron shows no preference for antiparallel ejection with respect to either of 
the fixed electrons. 
The ${\bf T}$-shape dominance seems to be well established at 10~eV above the (triple) ionization threshold (at least for the specific break-up configurations
so far analyzed), which is not quite in the Wannier threshold region (usually characterized as several eV above the threshold). 

In figure~2 we present more evidence for the dominance of the ${\bf T}$-shape, by looking at angular distributions with fixed electrons at $0^{\circ}$ and
$90^{\circ}$. This distribution is not identical to the distribution in figure~1, even though the relative angle between the fixed electrons is the same. This is due
to the influence of the polarization axis on the angular distribution, something that has long been recognized in the simpler two-electron double photoionization \cite{ah05}.
At large excess energies, we see that the angular distribution of the third electron is peaked around angles of 200--220$^{\circ}$, again close to the angle that
maximizes the third electron's distance from the fixed electrons. As the excess energy is decreased, we find that this peak moves towards an
ejection angle of $180^{\circ}$, a direction that is well established at excess energies of 10 and 5 eV and forms a clear ${\bf T}$-shape with the fixed
electrons. At these lower excess energies, we also find
a small peak along ejection angles of $270^{\circ}$. This smaller peak also corresponds to a ${\bf T}$-shape ejection, but the magnitude of the peak along this 
direction is much smaller than the peak along $180^{\circ}$, unlike the case examined in figure~1, where the two peaks were of identical size. The difference in this
case is due to the polarization direction of the radiation field; the ejection along the field (which is horizontal in all the plots shown here) is more probable
than ejection perpendicular to the polarization direction. Ejection along the field polarization direction is also more likely in (two-electron) double photoionization
of a $1s2s$ electron pair, as discussed previously \cite{cp03}.

\begin{figure}
\label{fig2}
\includegraphics[scale=0.4]{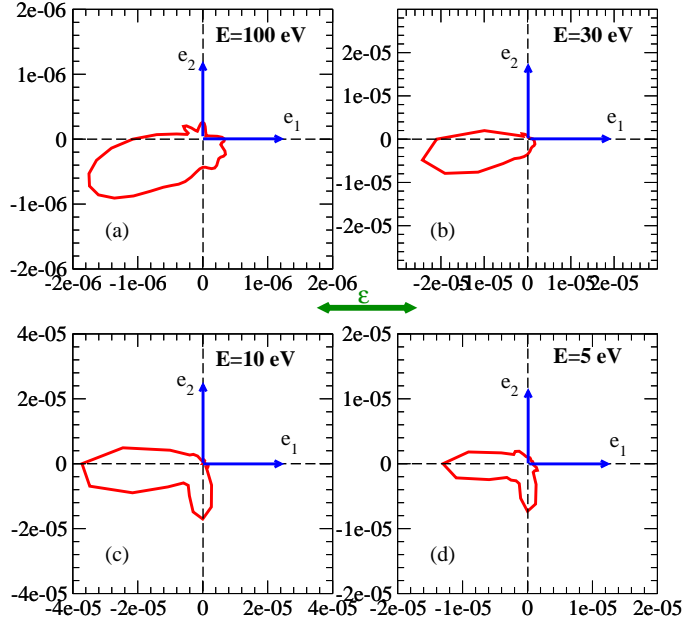}
\caption{(Color online) Pentuple differential cross sections for triple photoionization of Li at 
four excess energies (E) as indicated, $E_1=E_2=E_3$,  and for $\theta_1=0^{\circ}$, $\theta_2=90^{\circ}$. Results are presented as a 
function of $\theta_3$ and for coplanar geometry.
All cross sections are in units of b/(sr$^3$ eV$^2$).
}
\end{figure}

\begin{figure}
\label{fig4}
\includegraphics[scale=0.4]{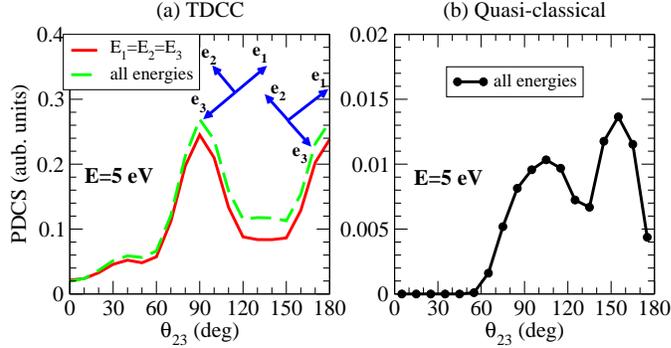}
\caption{(a) TDCC calculations of pentuple differential cross sections for triple photoionization of Li at 
5~eV excess energy, for fixed relative angle between electrons 1 and 2 of $\theta_{12}=90^{\circ}$, and as a function of the relative
angle $\theta_{23}$ between electrons 2 and 3, all in the coplanar geometry. The cross section is summed over all possible
$\theta_1$ values with respect to the polarization direction (defined by $\theta_1=0^{\circ}$). The red curve shows the equal energy
sharing cross section and the green dashed curve shows the cross section summed over all possible electron energy sharings.
(b) Quasi-classical calculations of the triple photoionization of Li where all electrons are ejected into a common plane. Only those events are 
included where the external product of two of the outgoing momenta vectors make an 
interelectronic angle between $87.5^{\circ}$ and $92.5^{\circ}$ with the third vector.
}
\end{figure}

As final evidence for the dominance of the ${\bf T}$-shape break-up, we present in figure~3 angular distributions that are for a fixed
relative angle between the two fixed electrons of $\theta_{12}=90^{\circ}$, in the coplanar geometry, 
as a function of the relative angle between electrons $2$ and $3$, at an excess energy of 5~eV. Fig.~3(a)
shows TDCC calculations that have been summed over all values of $\theta_1$ from $0$ to $180^{\circ}$, so that the influence of the polarization field on the
angular distribution is effectively averaged over. The figure shows the angular distribution for equal energy sharing case (solid red line) and for all
possible energy sharings (dashed line); both sets of calculations are quite similar. We find two prominent peaks in the distribution at 
$\theta_{23}=90^{\circ}$ and $\theta_{23}=180^{\circ}$; each of these corresponds to a ${\bf T}$-shape break-up as indicated.
The magnitude of the peaks at $90^{\circ}$ and $180^{\circ}$ are almost identical, showing that once the influence of the polarization direction is accounted for,
the third electron is equally likely to form a ${\bf T}$-shape by moving antiparallel to electrons $1$ or $2$. Several further calculations performed in non-coplanar
geometries (not shown) also find angular distributions that are similar to those shown in figure~3. We find that the ${\bf T}$-shape break-up pattern is also evident
for the angular distribution summed over all energy sharings (dashed line), allowing us to conclude that this break-up pattern is found at low excess energies irrespective of the 
energy sharing between the electrons, 
For such relative angle distributions, we may also compare with the predictions
of quasi-classical calculations \cite{er06,ewr08}. Figure~3(b) shows such a calculation where the electrons are ejected into a common plane (the polarization axis does not enter into
the quasi-classical calculations so this plane cannot be identified with the coplanar case shown in figure~3(a)), for any possible energy sharing between the electrons. The
quasi-classical prediction is also of two prominent peaks near relative angles of $90^{\circ}$ and $180^{\circ}$, although not at quite the same relative angles as predicted
in the TDCC calculations shown in figure~3(a). The difference between the TDCC quantal calculations and quasi-classical calculations may be due to the restriction made in the
TDCC calculation to only a coplanar geometry, or, it could be due to the neglect of the polarization axis influence within the quasi-classical approach. 
Convolution with the volume element $\sin\theta$
also leads to an appearance of the 180$^{\circ}$ peak at a smaller angle for the quasi-classical calculation.
A complete set of
TDCC calculations for all possible planes, which would be required for a direct comparison with the quasi-classical calculations \cite{er06,ewr08}, 
is a daunting task since it would require calculations for all possible relative angles for all three electrons, in all possible planes, 
and for all possible energy sharings. However, the good qualitative agreement between the TDCC calculations and quasi-classical predictions provides strong evidence for the
dominance of the ${\bf T}$-shape break-up at these excess energies.

Since we have now established that a ${\bf T}$-shape is the dominant break-up configuration at low excess energies, it is important to understand the physics behind
such a pattern, as it contrasts with the general expectation that the Wannier threshold break-up pattern  also holds a few eV above threshold, 
such as the 5 eV energy considered here.
Our picture of the triple photoionization process is the following. The photon is absorbed by one of the $1s$ electrons and immediately ionized. 
Cross sections for single ionization from the $1s$ sub-shell at photon energies above 200 eV are several orders of magnitude larger than the cross sections for
ionization from the $2s$ sub-shell. 
 After this rapid ionization, there are several mechanisms through which the remaining $1s2s$ electron pair can escape. Shake-off of the $2s$ electron after removal of
{\it both} $1s$ electrons was proposed in early studies of this process \cite{weh98,weh00}, and a double shake-off mechanism was postulated for 
the related process of triple
photoionization of Be \cite{kb03}. Both of these processes are expected to be significant at large photon energies, but not at the low excess energies currently considered.
Successive knock-out processes were also proposed \cite{er07}, whereby the photoelectron knocks out the second $1s$ electron and then the
remaining $2s$ electron as it leaves the atom. A similar process where the photoelectron knocks out the second $1s$ electron, and this second $1s$ electron removes the 
$2s$ electron was also considered. The signature of these break-up mechanisms should be reflected in the relative energies of the outgoing electrons; we
should expect to find that triple ionization events where one electron has most of the excess energy, and the remaining two have smaller (approximately equal)
energies, are more likely. This is indeed the case, as demonstrated in figure~4, where we show the angular distribution for fixed back-to-back electrons 
($\theta_1=90^{\circ}$; $\theta_2=270^{\circ}$) and where the energy of the third electron is increased from 33\% to 90\% of the available excess energy.  We find
the shape of the distribution to be a broad ${\bf T}$-shape in all cases, but the magnitude 
of the cross section increases by over a factor of four as the third electron retains more of the available excess energy. Further analysis of the electron energetics
at a fixed set of angles corresponding to the ${\bf T}$-shape reveals that the most probable configuration is where the third electron carries off most of the energy, and that
the energy sharing between the remaining electron pair is less important.
The $1s2s$ outgoing electron pair may have singlet or triplet characteristics, depending on their orientation with respect to the other $1s$
electron, but in either case, these
electrons are likely to escape in a back-to-back configuration, since this maximizes their mutual angle, and since this electron pair has a total angular momentum of 0. 
Electron repulsion will also position this electron pair as far as
possible from the direction of the first electron: this leads naturally to a ${\bf T}$-shape break-up configuration. 
This conclusion was also previously reached in \cite{ewr08} where, after assuming that the $1s2s$ electron pair breaks up back-to-back, the ${\bf T}$-shape break-up
configuration was shown to be the most stable configuration.
We finally note that consideration of the selection rules for two-electron ejection \cite{mb95} that are very instructive in analyzing
double photoionization angular distributions do not shed more light 
on the expected ejection patterns of the residual $1s2s$ electron pair. This is because the two electrons are in different subshells and because
the overall symmetry of this $1s2s$ electron pair is $L=0$ (since the photoelectron absorbs the photon and thus has the available one unit of angular momentum).
The dominant emission pattern for the break-up of these two electrons is simply back-to-back, as we show.

\begin{figure}
\label{fig5}
\includegraphics[scale=0.4]{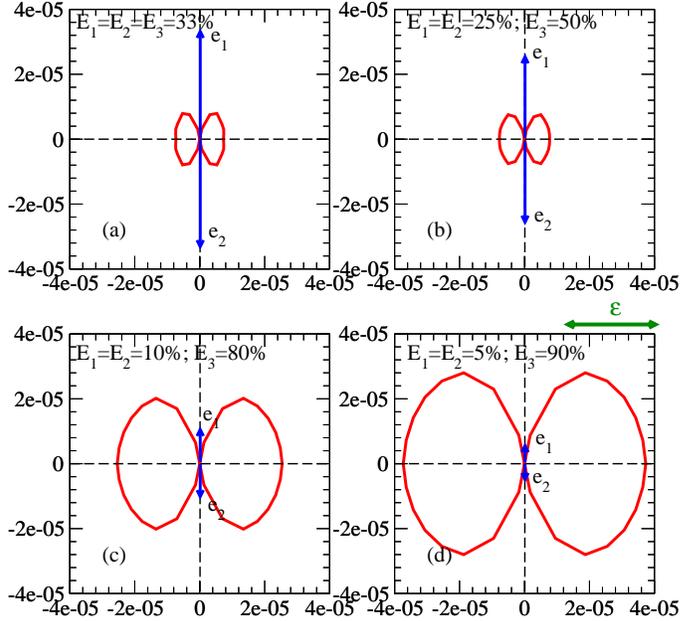}
\caption{(Color online) Pentuple differential cross sections for triple photoionization of Li at 
an excess energy of 5 eV and for $\theta_1=90^{\circ}$, $\theta_2=270^{\circ}$. Results are presented as a 
function of $\theta_3$, for coplanar geometry and for various excess energies as indicated.
All cross sections are in units of b/(sr$^3$ eV$^2$).
}
\end{figure}

Therefore, the ${\bf T}$-shape break-up dynamic can arise from the rapid removal of a $1s$ electron followed by back-to-back ejection of the remaining electron pair. 
The ${\bf T}$-shape formation results from the minimization of the energy of this configuration \cite{ewr08}, which leads to the back-to-back electron pair oriented at 90$^{\circ}$ to
the fast electron.
At larger excess energies the ${\bf T}$-shape break-up is also viable, but becomes one of several competing break-up configurations \cite{cp12}, resulting in more 
complicated angular distributions.

The Los Alamos National Laboratory is operated by Los Alamos
National Security, LLC for the NNSA of the U.S. DOE under Contract No.
DE-AC5206NA25396.
This work was supported in part by grants from the U.S. DOE and the U.S. NSF to Auburn University.
Computational work was carried out at the NERSC in Oakland, California,
and at Los Alamos National Laboratory.

\end{document}